\documentclass{JINST}

\usepackage{amsmath}

\newcommand{\ud}{\,\mathrm{d}}

\title{Luminosity measurement at ILC}

\author{I. Bo\v zovi\'c Jelisav\v ci\' c$^a$, S. Luki\'c$^a$, G. Milutinovi\'c Dumbelovi\'c$^a$, M. Pandurovi\'c$^a$ and I. Smiljani\'c$^a$\\
\llap{$^a$}Vin\v{c}a Institute of Nuclear Sciences, University of Belgrade\\
M. Petrovi\'ca Alasa 12-14, 11001 Belgrade, Serbia\\
  E-mail: \email{ibozovic@vinca.rs}}

\abstract{In this paper we describe a method of luminosity measurement at the future linear collider ILC that estimates and corrects for the impact of the dominant sources of systematic uncertainty originating from the beam-induced
effects and the background from physics processes.  
Based on the relativistic kinematics of the collision frame of the Bhabha process, the beam-beam related uncertainty is reduced to a permille independently of the precision with which the beam parameters are known. With the specific event selection, different from the isolation cuts based on topology of the signal used at LEP, combined with the corrective methods we introduce, the overall systematic uncertainty in the peak region above 80\% of the nominal center-of-mass energy meets the physics requirements to be at the few permille level at all ILC energies.}


\begin{document}

\section{Introduction}

Integrated luminosity measurement at a future linear collider will be performed by counting Bhabha events reconstructed
in the luminometer fiducial volume (FV) within specified event selection. To match the physics benchmarks (i.e. W-pair
production,  fermion pair-production, cross-section measurements) that might be of particular interest for the new
physics, luminosity should be known at the level of $10^{-3}$ or better \cite{ILD10}. In this paper we present the method
of measurement optimized to meet the requirements for luminosity precision. Using MC simulation of the relevant physics and beam-induced processes, we show that the luminosity can be
measured in the peak region above 80\% of the nominal center-of-mass (CM) energy with a relative precision of a few permille
at all ILC energies. In addition, luminosity spectrum can be precisely reconstructed in the same energy region from the
experimentally measurable quantities.

Finely granulated calorimeters, of high energy and polar angle resolution, are foreseen to instrument the very forward
region at ILC. Luminometer at a future linear collider will be designed as a compact sampling calorimeter with Moliere
radius of approximately 1.5 cm \cite{JINST2010}. To reduce systematic biases from the mechanical precision of the alignment, a
laser based monitoring system has been developed for ILC to control the position of the luminometer over short distances
within a micron \cite{Aguilar}. ILD detector model \cite{ILD10} for ILC is assumed at CM energies of 500
GeV and 1 TeV. Design of the luminometer is described in Section \ref{sec:Lumi}. Event simulation used to establish the presented method of luminosity measurement is described in Section \ref{sec-sim}.

With the rising energy and the bunch density, one of the main uncertainties in luminosity measurement at a future
$e^{+}e^{-}$ collider at TeV energies comes from the effects induced by space charges of the opposite beams.
Beamstrahlung and electromagnetic deflection induced by the field of the opposite bunch, together with the initial state
radiation (ISR), result in the change of the four-vectors of the initial and final state particles, consequently causing
the deviation of the polar angles and counting losses of the signal in the luminometer. Dominating beamstrahlung effects
are particularly pronounced at higher CM energies, resulting in 12.8\% counting loss at 500 GeV at ILC \cite{LCnote}, assuming beam parameters from Ref. \cite{IDR11} and 18\% counting loss in the 3 TeV CLIC case \cite{Luk13}, in the upper 20\% of the luminosity spectrum. If the full energy range is considered, counting losses are more severe ranging up to the 70\% at 3 TeV CLIC. Corrective methods developed to take this into account are
reviewed in this paper for the 500 GeV and 1 TeV ILC cases in Sections \ref{sec:coll} and \ref{sec:EMD}.

Due to the precision requirements of luminosity measurement, background from physics processes is discussed as another important source of systematic uncertainty. Electron spectators from the four-fermion processes can be misidentified as signal since they are emitted at low angles and with high energy. These processes can be, in principle, efficiently suppressed by the LEP type isolation cuts based on the signal topology \cite{Mila00, MI} that can not be directly translated to the linear collider case where beam-induced effects have to be simultaneously taken into account. Physics background in luminosity measurement is discussed in detail in Section \ref{sec:4f}.

Summary on systematic uncertainties in luminosity measurement is given in Section \ref{syst_sum}.

\section{Luminometer at ILC}
\label{sec:Lumi}

\begin{figure}[htp]
\begin{center}
\includegraphics[width=.45\textwidth]{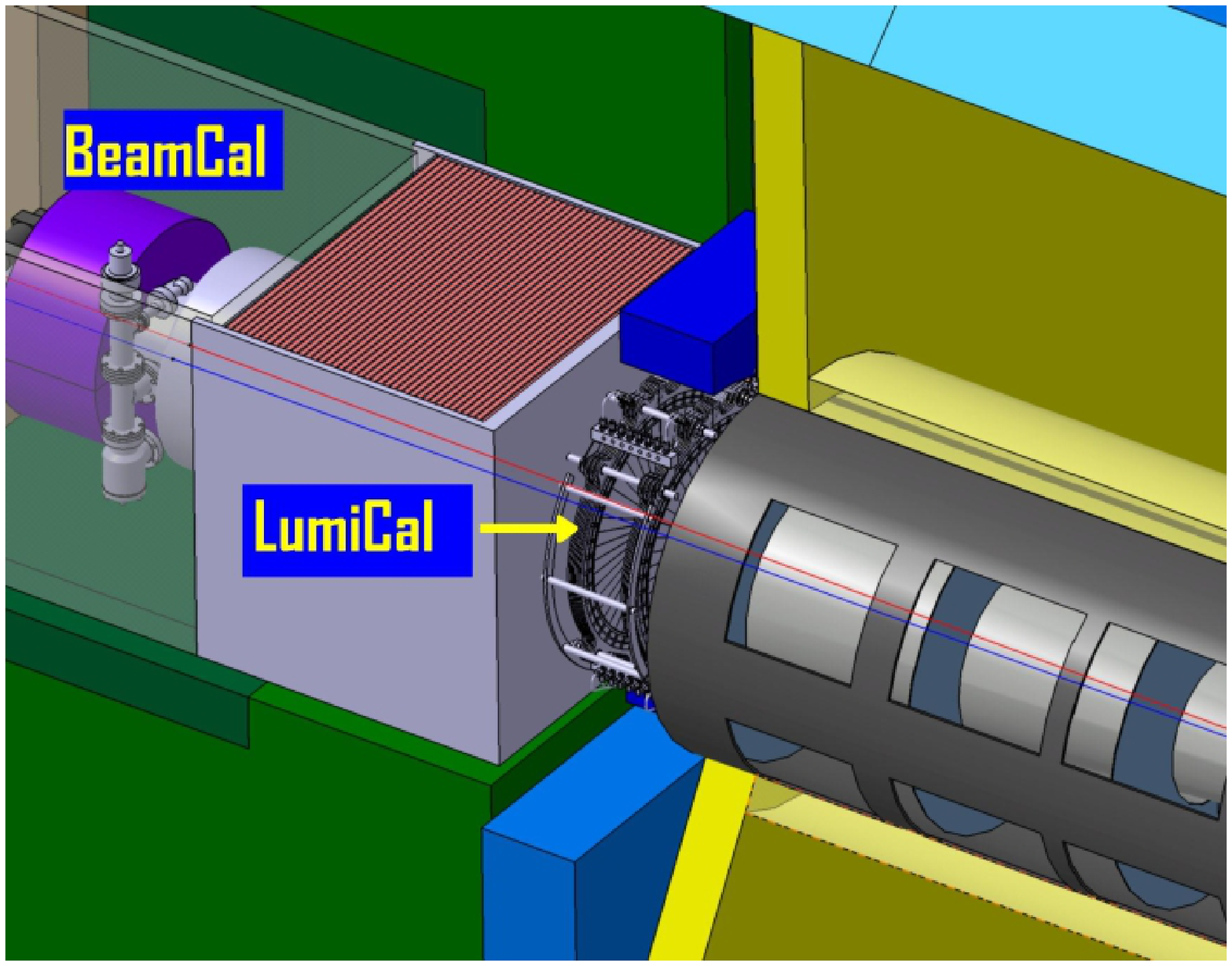}
\includegraphics[width=.334\textwidth]{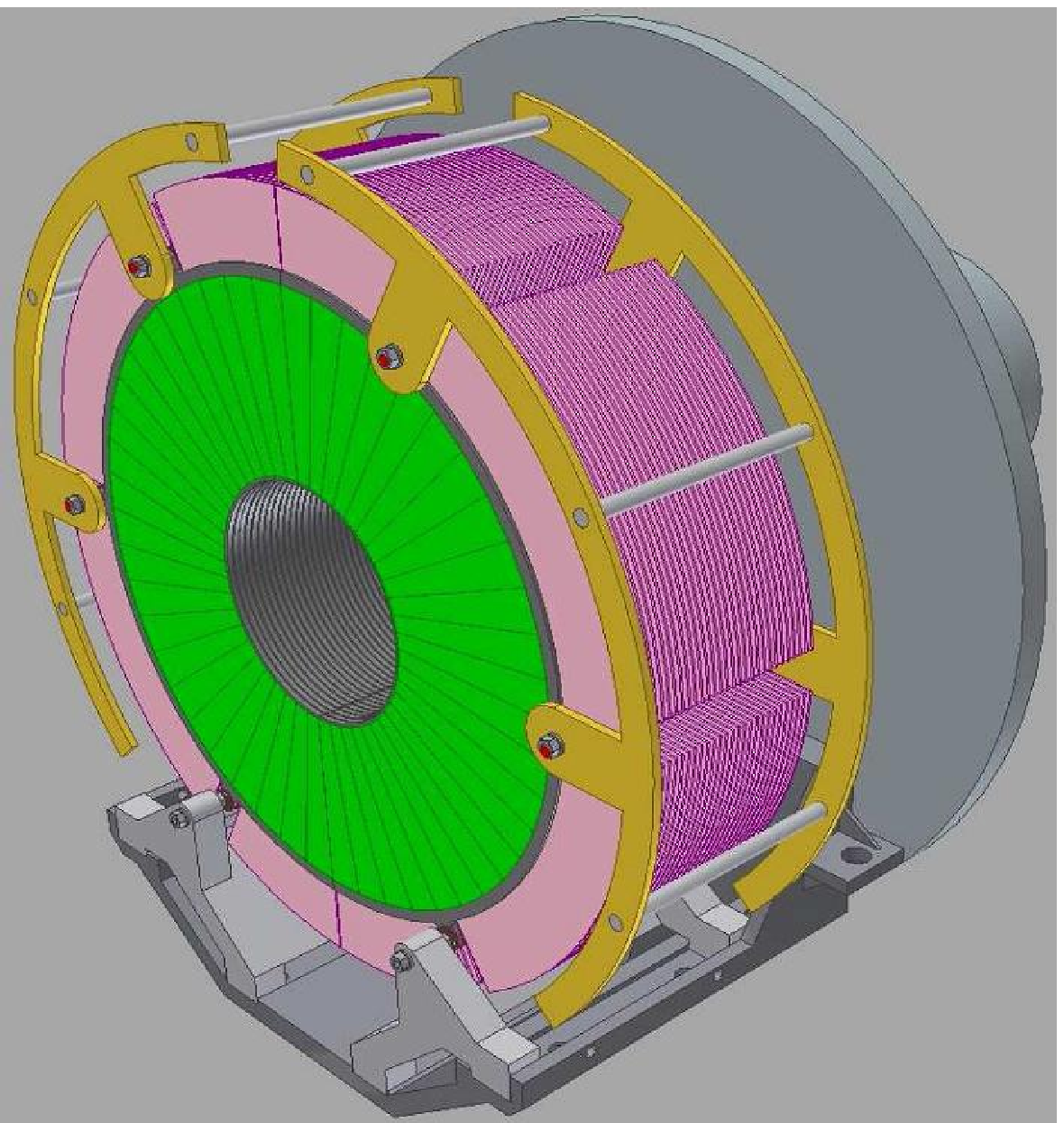}
\caption{\label{LCAL}Layout of the very forward region of ILD (left). Mechanical structure of the luminosity calorimeter
for ILD (right).}
\end{center}
\end{figure}

Two concepts of particle detectors are being developed for ILC, the International Large Detector (ILD) \cite{ILD10} featuring a Time Projection Chamber as the central tracking system and the Silicon Detector (SiD) \cite{SiD09} with a compact semiconductor central tracker designed to optimize the physics performance, as well as the cost. Both detectors have similar layout, combining excellent tracking and finely-grained calorimetry, which enables energy reconstruction of individual particles using particle flow algorithm \cite{PF}. To provide jet separation in multi-jet processes, electron identification down to the lowest polar angles, as well as a good hermeticity, the very forward region is instrumented with the two special calorimeters: luminometer designed to measure the rate of the low angle Bhabha scattering and the beam calorimeter that will allow fast luminosity estimate and measurement of the beam parameters. Layout of the very forward region of the ILD detector is illustrated in Figure \ref{LCAL} (left) \cite{ILD10}. Both the ILD and SiD concepts share the same design of the luminometer, the only difference being in the relative longitudinal positioning of the luminometer with respect to the IP. The present study refers to the ILD geometry.

The luminometer itself is foreseen as a sampling silicon/tungsten calorimeter, consisting of 30 absorber planes, each with thickness of one radiation length (3.5
mm), interspersed by segmented silicon sensor planes. To keep the Moliere radius of 1.5 cm, sensor gaps are kept at 1 mm. As illustrated in Figure \ref{LCAL} (right), tungsten disks are precisely positioned using 4 bolts. The system is additionally stabilized by steel rings. The accuracy of the electron polar angle reconstruction depends on the sensor segmentation. The optimized layout contains 48 azimuthal and 64 radial divisions, yielding a predicted angular resolution of $\sigma_{\theta}=(2.20\pm0.01) \cdotp 10^{-2}$ mrad and the polar angle bias $\Delta\theta=(3.2\pm0.1) \cdotp 10^{-3}$ mrad \cite{JINST2010}. Polar angle measurement bias, due to the clustering algorithm \cite{Sad08}, is defined as the central value of the difference between the reconstructed and the true value of the polar angle. Non-zero values of the polar angle bias are induced by the non-linear signal sharing of finite size pads with gaps between them. The contribution of each of these two uncertainties to the relative uncertainty of luminosity measurement is $1.6 \cdotp 10^{-4}$ \cite{JINST2010}. In the ILD version, the luminometer is positioned at 2.5 m from the IP, with the geometrical aperture between 31 mrad and 78 mrad. Inner and outer radius are 80 mm  and 195.2 mm respectively. Since the cross-section for Bhabha scattering is falling with the polar angle as $1/\theta^{3}$, the inner aperture of the luminometer has to be known below the permille \cite{Stahl05} to keep the counting
uncertainty below permille. The constraint on relative lateral position of the IP with respect to the LumiCal is much less strict because of the averaging over the azimuthal angle. An uncertainty of several hundred microns in the relative lateral IP position induces an uncertainty of $10^{-4}$ on the luminosity measurement \cite{Stahl05}.

With 30 radiation lengths of tungsten as the absorber, high energy electrons and photons deposit almost all of their energy
in the detector. The relative energy resolution $\sigma_{E}/E$ is parametrized as:

\begin{equation}
{\frac{\sigma_{E}}{E}=\frac{a_{res}}{\surd{E_{beam}(GeV)}},}
\end{equation}

where $E$ and $\sigma_{E}$ are, respectively, the central value and the standard deviation of the distribution of the
energy deposited in the sensors for a beam of electrons with energy $E_{beam}$. The sampling parameter $a_{res}$ is usually
quoted as the energy resolution and it is estimated to be $a_{res}=(0.21\pm0.02)$ $\surd{GeV}$ \cite{JINST2010} for
electron showers located inside the FV of the luminometer. The detector FV, where $\sigma_{E}/E$ is
practically constant over $\theta$, extends from 41 to 67 mrad.

\section{Event simulation}
\label{sec-sim}
\subsection{Simulation of the signal influenced by the beam-induced effects}
\label{sec-sim-bh}

To simulate the influence of the beam-induced effects on signal, Guinea-Pig software 1.4.4 \cite{Sch96} was used to simulate the collisions of bunches. At the point when the initial four-momenta of the colliding $e^{-}e^{+}$ pairs are generated, the decision is made by Guinea-Pig whether the Bhabha scattering will be realized in the collision. The decision is made randomly, based on the cross-section for the Bhabha scattering at the CM energy of the colliding pair. If Bhabha event is to be realized, the final four-momenta are picked from a file generated at the nominal ILC
CM energy (500 GeV, 1 TeV) with the BHLUMI V4.04 generator  \cite{Jad97}. 
The BHLUMI generator was run with the cuts on the polar angles in the laboratory frame $\theta_{min}^{lab} = 10 \text{\, mrad}$ and $\theta_{max}^{lab} = 200 \text{\, mrad}$. After generation, post-generator cuts were applied on the scattering angle $\theta^{coll}$ in the collision frame, and only events with $37 \text{\, mrad} < \theta^{coll} < 75 \text{\, mrad}$ were kept in the event file to be read by the Guinea-Pig. The cuts on the scattering angle are somewhat wider than the angular range of the luminometer FV to ensure a safety margin for the angular shift due to EMD and off-axis radiation. On the other hand, the cuts in the lab frame are relaxed because the longitudinal boost due to beamstrahlung requires a much wider margin. The momenta from the generator file are then
scaled to the CM energy of the colliding pair, rotated to match the collision axis and then boosted back to
the laboratory frame. Finally, electromagnetic deflection of the final state is simulated using the Guinea-Pig feature
to predict the final deflection angles.  

The standard beam-parameter set from the ILC Technical Progress Report 2011 \cite{IDR11} was used as the basis for both 500
GeV and 1 TeV simulations. Variations of individual beam parameters are taken into account in order to determine the
influence of the beam-parameter uncertainties on the performance of the presented method (Section \ref{sec:coll}). Simulated
beam-parameter variations include:

\begin{itemize}
 \item Symmetric bunch size variations by $\pm 10$ and $\pm 20\%$ and one-sided variations by +20\% in $\sigma_{x}$,
 $\sigma_{y}$, $\sigma_{z}$;
 \item Symmetric bunch charge variations by $\pm 10$ and $\pm 20\%$ and one-sided +20\% variation;
 \item Relative misalignment of the two beams in the x- and y-directions by up to one respective bunch RMS width. 
\end{itemize}

Thus, 25 sets of beam parameters were simulated in total for each of the two ILC energy options. In each simulation, one single beam parameter was varied with respect to its nominal value. Between 1.5 and 4 million Bhabha events were generated in each simulation. The interaction with the detector was approximated in the following way:

\begin{itemize}
 \item The four-momenta of all electrons and photons, that are treated as indistinguishable, are summed together in the 5 mrad
 cone around the most energetic shower. The 5 mrad criterion corresponds closely to the Moli\`ere radius of the high-energy showers 
 in the LumiCal \cite{Sad08}. The initial beamstrahlung photons were not included as they are emitted close to the beam axis. For
 synchrotron radiation, the characteristic emission angles are of the order $1/\gamma$, being smaller than $10^{-3}$ mrad
 for electron energies in the TeV range, therefore the four-momenta of photons emitted by the outgoing electrons can be
 added. 
  \item The energy resolution of the luminometer is simulated by adding random fluctuations to the final particle energies.
 The random fluctuations were sampled from the Gaussian distribution with the energy-dependent standard deviation. 
 \item The finite angular resolution of the luminometer is included by adding random fluctuations to the final particle
 polar angles corresponding to the  luminometer resolution in polar angle as given in Section 2.
\end{itemize}

\subsection{Simulation of physics background}
\label{sec-sim-bkgd}

Four-fermion events  $e^{+} e^{-}\rightarrow e^{+} e^{-} f\bar f$ are generated at tree level at 500 GeV and 1 TeV
CM energy, with the total corresponding cross-section of  $\sigma_{bck}=(5.1\pm0.1)$ nb and 
$\sigma_{bck}=(0.8\pm0.1)$ nb respectively,
using WHIZARD-V1.2 event generator \cite{Whiz}. Parameters of the event generation are tuned using $e^{+} e^{-}\rightarrow
e^{+} e^{-} c \bar c$  to describe the experimental results obtained by several experiments at lower energy at PETRA and
LEP accelerators \cite{Pozdnyakov}. Thus the physics background is generated under the following assumptions:

\begin{itemize}
 \item Mandelstam invariant mass of the outgoing lepton pair is greater than 1 GeV$^{2}$,
 \item Momentum transfered by the photon exchange is greater then $1\cdot10^{-4}$ GeV,
 \item Events were generated within the polar angle range between 0.05 and 179.95 degrees to avoid the divergence of the
 cross-section  at  low polar angles.
\end{itemize}

Matrix elements for the leading order Feynman diagrams are computed using O'Mega generator \cite{Omega}.  The total sizes of
samples correspond approximately to 1 million of background events integrated over the polar angle range at each ILC energy. The cross section for
background processes integrated in the polar angle range of the luminometer FV is approximately a percent of the
generated one.

Influence of physics background is estimated against the signal samples of 20 pb$^{-1}$ of Bhabha events, with the
cross-sections in the luminometer fiducial volume of $\sigma_{s}$ = (4.689$\pm$ 0.001) nb and
$\sigma_{s}$ = (1.197$\pm$0.005) nb at 500 GeV and 1 TeV CM energy, respectively. Estimated errors of the
cross-sections are statistical. Signal is simulated with the BHLUMI event generator \cite{Jad97} implemented in BARBIE V5.0
\cite{Barbie}, a GEANT4 \cite{Geant4} based detector simulation of the luminometer at ILC. 

\section{Luminosity measurement}

\subsection{Method of luminosity measurement}
\label{sec:method}

Luminosity at electron-positron collider is measured using Bhabha scattering at low angles ($e^{+}e^{-} \rightarrow e^{+}e^{-}(\gamma)$) as a gauge process, which is calculable with excellent precision in QED. This is dominantly electromagnetic
process (99\% at ILC energies \cite{Jad02}) of one-photon exchange in the t-channel for which the cross-section calculations
with relative uncertainty better than $10^{-3}$ are available \cite{Jad03}. Luminosity at ILC will be measured by counting
Bhabha scattering events that are recognized by coincident detection of showers in the two halves of luminometer. Unless
distorted by the beam-induced effects, these showers are collinear and carry the largest fraction of the available CM energy.
The luminosity integrated over certain period of time is calculable as the number of Bhabha events, $N$, divided by the
theoretical cross-section for the Bhabha scattering, $\sigma$.

\begin{equation}
\label{eq-luminosity_0}
L = \frac{N}{\sigma}
\end{equation} 

In the most general view, equation (4.1) translates to:

\begin{equation}
\label{eq-luminosity}
L = \frac{N(\Xi(\Omega^{lab}_{1,2}, E^{lab}_{1,2}))}{\sigma(Z(\Omega^{CM}_{1,2}, E^{CM}_{1,2}))}
\end{equation}

Here $\Omega$, $E$ stands for angular parameters and energy of the final state particles; the superscript specifies whether these quantities are evaluated in the laboratory or the CM frame; and  $\Xi$ (resp. $Z$) represent the ensemble of selection criteria and kinematic cuts applied to the event selection (resp. the cross-section evaluation). The notation emphasizes that $\Xi$ and $Z$ operate on kinematical arguments in different reference frames. The cross section is typically calculated by the Monte Carlo integration using an event generator in the CM frame of the process, under assumptions expressed by $Z$. On the other hand, Bhabha particles are experimentally reconstructed in the laboratory frame under certain selection criteria $\Xi$. In addition, the final state four-momenta are affected by the beam-beam
effects. Because of random and asymmetric emission of the beamstrahlung and ISR, the CM frame of the Bhabha
process is different for every colliding pair and in general it does not coincide with the laboratory frame. The
resulting axial boost of the outgoing particle angles induces angular acceptance loss of the signal in the detector
FV. An additional systematic bias of the order of 1 to 2 permille \cite{LCnote, Rim07} is induced by the
electromagnetic deflection (EMD) of the outgoing Bhabha particles in the field of the opposite bunches.

At LEP, angular counting losses were usually addressed by applying selection in polar angles of the final state particles that is
asymmetric  with respect to the forward and backward side of the luminometer \cite{Opal00, Aleph00, L3_00}. In order to minimize the beam induced effects on Bhabha counting rate, similar techniques were proposed for ILC as well \cite{Stahl05, Rim07} where the beam-beam effects are much more intense. The beam-induced bias minimized in this way is then corrected by application of beam-beam simulation to determine the correlation between the bias and the form of the reconstructed luminosity spectrum. The above mentioned corrective techniques are simulation dependent and the systematic uncertainty of luminosity measurement depends on the knowledge of the beam parameters (i.e bunch sizes) \cite{Rim07}.

To overcome these limitations, a concept introduced in Ref. \cite{Luk13} is used, in which one defines $\Xi$ and $Z$ in such a way that the counting rate is independent of reference frames in which these functions act. In the following section we briefly review this method and test it in the ILC case using MC simulation. The EMD effect requires a different approach that will be discussed in Section 4.1.2. In addition, event selection has to be established in a way to suppress the background from physics processes present in the very forward region (Section \ref{sec:4f}).

\subsubsection{Collision-frame method}
\label{sec:coll}

The collision frame of a scattering reaction is defined here as the CM frame of the electron-positron system after emission of ISR and before the emission of FSR. In the collision frame the deflection angles of both colliding particles are the same due to the momentum conservation principle. The movement of the collision frame with respect to the laboratory frame is responsible for the acollinearity leading to the angular counting loss. Since the kinematics of the detected showers correspond to the system after ISR, the velocity of the collision frame $\vec{\beta} _{coll}$ can be calculated from the measured polar angles. Since Beamstrahlung and ISR are approximately collinear with the incoming electrons, $\vec\beta_{coll}$ can be taken to be approximately collinear with the $z$-axis. Under this approximation, the modulus of $\vec\beta_{coll}$ can be expressed as:

\begin{equation}
\label{eq-beta}
\beta_{coll} = \frac{sin(\theta^{lab}_{1}+ \theta^{lab}_{2})}{sin(\theta^{lab}_{1})+ sin(\theta^{lab}_{2})}
\end{equation}

\begin{figure}
\centering
\includegraphics[width=.4\textwidth]{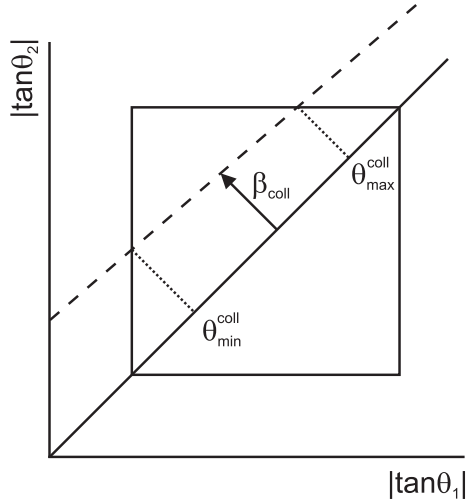}
\caption{\label{fig-beta_coll} Schematic representation of the distortion of the polar angles due to the movement of the collision frame. The box represents the region in which both electrons hit the FV, and the dashed line represents the event subset characterized by a given $\beta_{coll}$. $\theta^{coll}_{\min}$ and $\theta^{coll}_{\max}$ denote the effective limiting scattering angles for this subset.}
\end{figure}

Due to the longitudinal boost of the final particle angles, the effective acceptance of Bhabha events in the luminometer decreases with increasing $\beta_{coll}$ (see Figure \ref{fig-beta_coll}). The effective limiting scattering angles $\theta^{coll}_{min}$ and $\theta^{coll}_{max}$ in the collision frame for a given $\beta_{coll}$ are obtained by boosting $\theta_{min}$ and $\theta_{max}$ into the collision frame. This allows calculating the event-by-event weighting factor $w(\beta_{coll})$ to compensate for the loss of acceptance:

\begin{equation}
\label{eq-w}
w(\beta_{coll}) = \frac{\int\limits^{\theta_{max}}_{\theta_{min}} \frac{\ud\sigma}{\ud\theta} \ud\theta }{\int\limits^{\theta^{coll}_{max}}_{\theta^{coll}_{min}} \frac{\ud\sigma}{\ud\theta} \ud\theta}. 
\end{equation}

For $\beta_{coll}$ larger than some critical value $\beta^{*}$ (approximately 0.24 in the ILC case \cite{ICHEP12}), the effective angular acceptance is zero and such events are irreducibly lost. In all other cases, the counting loss can be corrected using Equation \ref{eq-w}.

\begin{figure}
\centering
\includegraphics[width=1.\textwidth]{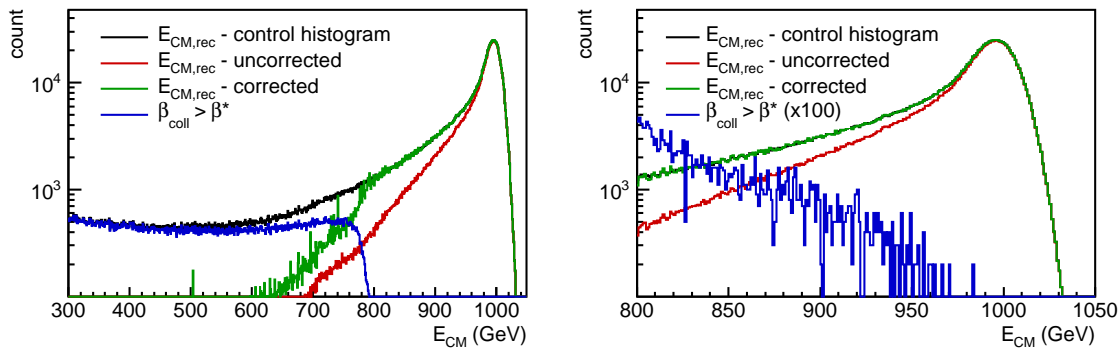}
\caption{\label{fig-BS-corr} Correction of the signal counting loss due to beamstrahlung and ISR at 1 TeV ILC (left). On the right, the same distributions are zoomed above 80\% of the nominal CM energy.}
\end{figure}

This correction method was tested using the simulation described in Section \ref{sec-sim-bh}. The performance of the method is tested by comparison of the corrected CM energy spectrum of Bhabha-events to the \emph{control} spectrum obtained by event selection based on the scattering angle in the collision frame, so the control spectrum corresponds to the signal events as if they were unaffected by the counting loss due to the longitudinal boost caused by Beamstrahlung and ISR. The results are shown in Figure \ref{fig-BS-corr} for the 1 TeV case. The control spectrum is plotted in black, the spectrum affected by the counting loss in red, the corrected spectrum in green. The blue line represents the events for which $\beta_{coll} > \beta^{*}$.

As can be seen from Figure \ref{fig-BS-corr}, despite severe counting losses due to beamstrahlung and ISR, the agreement after correction is excellent above 80\% of the nominal CM energy (800 GeV). The range below 80\% of the nominal CM energy is dominated by events for which $\beta_{coll} > \beta^{*}$. Due to kinematic constraints, high values of $\beta_{coll}$ are possible only with high energy loss, which explains the sudden drop of such events at 80\% of the nominal CM energy. However, a small number of events with apparent $\beta_{coll} > \beta^{*}$ is present also at energies above 80\% of the nominal CM energy, because occasionally the assumption that $\vec{\beta} _{coll}$ is collinear with the beam axis is broken due to off-axis ISR. 

The following is the list of sources of systematic uncertainty of the presented correction method:

\begin{enumerate}
\item\label{unc-off-axis} Off-axis ISR. In rare events with significant off-axis ISR, the assumption that $\vec{\beta} _{coll}$ is collinear with the beam axis does not hold, 
\item\label{unc-mes} The implicit assumption that the cluster around the most energetic shower always contains the Bhabha electron. In a small fraction of events of order of a few permille, this is not the case and the reconstructed polar angles $\theta_{1,2}^{lab}$ may differ from the final electron angles.
\item The use of the approximate angular differential cross section for the Bhabha scattering in the calculation of
$w(\beta_{coll})$,  
\item Assumption in the calculation of $\beta_{coll}$ and $w(\beta_{coll})$ that all ISR is lost and all FSR is detected.
\end{enumerate}

The relative bias due to the off-axis ISR is $-1.5 \cdotp 10^{-3}$ in the 500 GeV case and $-1.4 \cdotp 10^{-3}$ in the 1 TeV case. This bias is related to the energy and angular distributions of the ISR, which is reliably predicted by the generator. Thus this bias can be reliably corrected, and it is not sensitive to beam-parameter variations. In all simulations with the beam-parameter variations described in Section \ref{sec-sim-bh}, the bias due to off-axis ISR is constant to within 0.1 permille. 

The uncertainty introduced by the implicit assumption that the cluster around the most energetic shower always contains the Bhabha electron depends on the beam parameters and it may even depend on the specifics of the position-reconstruction algorithm in the luminometer. Therefore its correction will not be attempted here. The contribution of the effects 3 and 4 is smaller than the statistical uncertainty of the present analysis. The final quoted uncertainty will thus contain the contributions from the effects 2, 3 and 4.

In the following we present the numerical results of the correction method in terms of the fractional difference of the integral of the corrected to the control spectrum above 80\% of the nominal CM energy. In all individual simulations with beam-parameter variations described in Section \ref{sec-sim-bh}, the fractional difference is compatible with zero within the statistical uncertainty of the simulation, which is below permille in all individual simulations. For the overall precision, we quote the average over the entire set of simulations, for both energy options separately. Before correction of the bias due to off-axis ISR, the average fractional difference, containing contributions from all four systematic effects quoted above is $(-1.1\pm 0.1) \cdotp 10^{-3}$ in the 500 GeV case and $(-0.7\pm 0.1) \cdotp 10^{-3}$ in the 1 TeV case. After correction for the off-axis ISR, the remaining average fractional difference due to systematic effects number 2, 3 and 4 is $(+0.4 \pm 0.1) \cdotp 10^{-3}$ at 500 GeV and $(+0.7 \pm 0.1) \cdotp 10^{-3}$ at 1 TeV. The absolute size of these final biases can be taken as the present estimate of the uncertainty of the luminosity measurement induced by beamstrahlung and ISR.

One should note that the collision-frame method provides an accurate, simulation-independent correction of the angular counting loss of Bhabha events in the upper 20\% of energy spectrum. It does not correct for the beamstrahlung induced counting losses below 80\% of the nominal CM energy. The angular counting loss is reduced from ca. 10\% to ca. one permille in a simulation-independent way. Part of the remaining bias, due to off-axis ISR, is corrected by using the MC simulation result.

\subsubsection{Electromagnetic deflection}

\label{sec:EMD}

Electromagnetic field of the opposite bunch is causing shift of the outgoing particles toward smaller polar angles.
This shift is rather small, but since the Bhabha cross section is monotonously decreasing with the polar angle, the
net effect of EMD is a decrease of the Bhabha count. This effect is equivalent to the effect of a parallel shift in the angular limits $\theta_{min}$ and $\theta_{max}$ of the FV in the opposite direction by an \emph{effective mean deflection} angle $\Delta \theta$. 

\begin{equation}
\label{eq-emd}
\frac{\Delta L_{EMD}}{L} = \frac{1}{N} \frac{dN}{d\theta} \Delta \theta = x_{EMD} \Delta \theta
\end{equation}

The proportionality coefficient $x_{EMD} = \frac{1}{N} \frac{dN}{d\theta}$ between $\Delta\theta$ and the fractional counting loss $\Delta L_{EMD}/L$ is a quantity directly accessible in the analysis of either experimental or simulated data. On the other hand, to estimate $\Delta\theta$ beam-beam simulation has to be employed. 

\begin{figure}
\centering
\includegraphics[width=70mm]{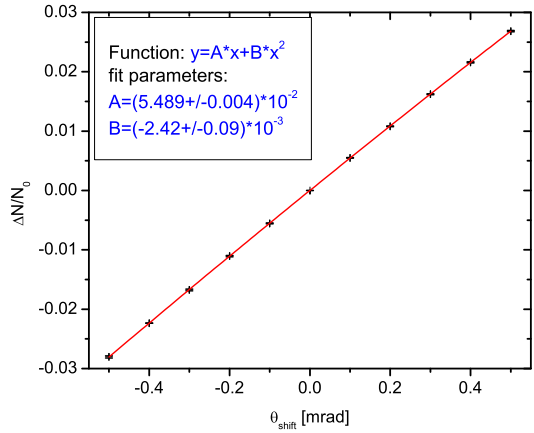}
\caption{\label{fig-dNdth}Fit of the $x_{EMD,sim}$ for the standard set of ILC beam parameters at 1 TeV.}
\end{figure}

Figure \ref{fig-dNdth} shows the fit of $x_{EMD,sim}$ for the standard set of beam parameters at 1 TeV. $\Delta N/N = (N_{shift} - N_{FV}) / N_{FV}$ is the fractional difference in counts in the shifted counting volume $(\theta_{min} + \theta_{shift}, \theta_{max} + \theta_{shift})$ versus the FV $(\theta_{min}, \theta_{max})$. The parameter $x_{EMD}$ is obtained as the slope of $\Delta N/N$ at the zero shift angle. The result in the simulation for the standard beam parameters is $x_{EMD} = 0.055 \text{\ mrad}^{-1}$. In the same simulation, the value of the fractional counting bias was $\Delta L_{EMD} /L = −1.07 \cdotp 10^{-3}$, which gives the resulting effective mean deflection angle $\Delta \theta = 0.020 \text{\ mrad}$. In the same manner, $\Delta \theta$ obtained at 500 GeV CM energy is $\Delta \theta= 0.043 \text{\ mrad}$.

\begin{figure}
\centering
\includegraphics[width=70mm]{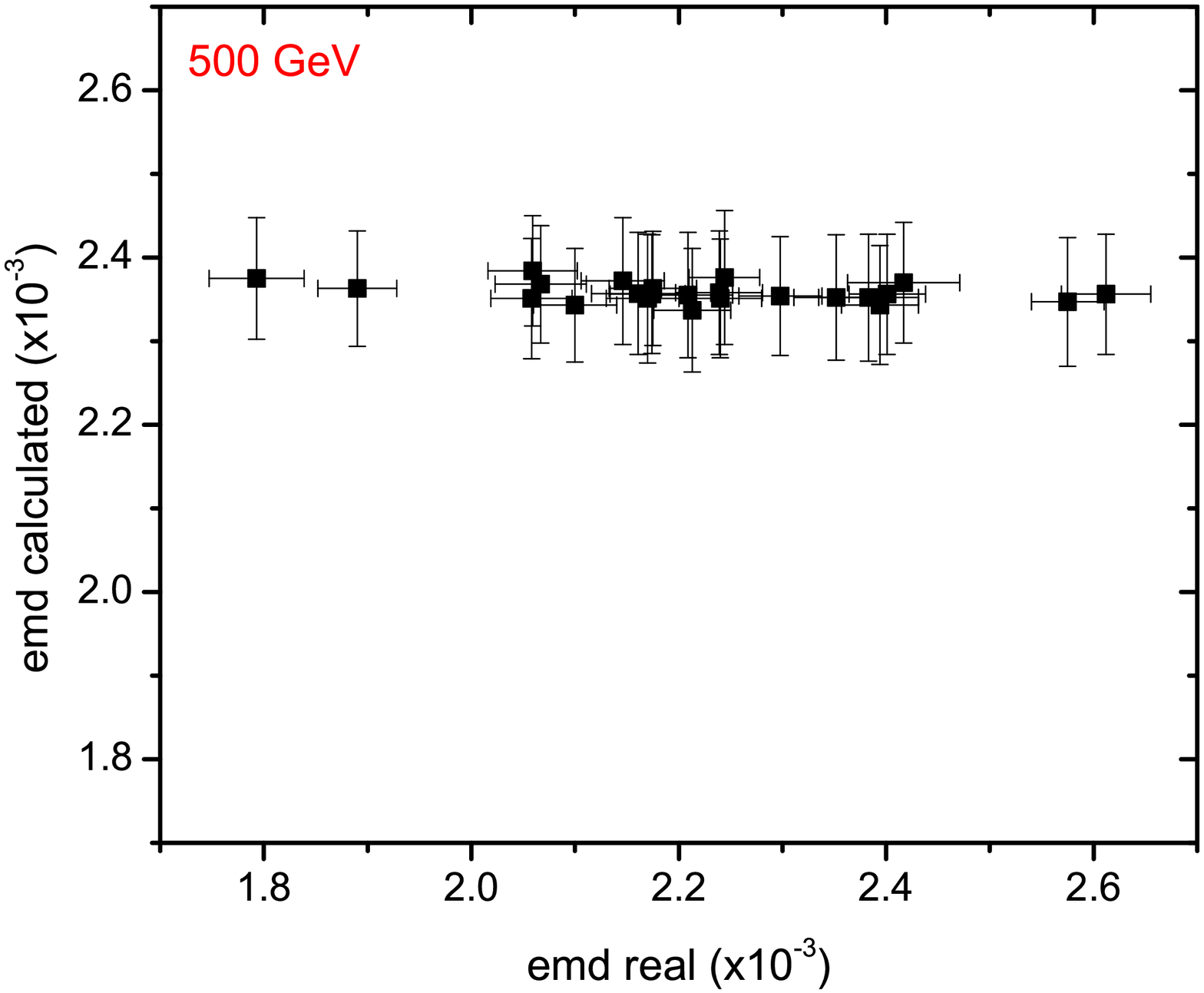}
\includegraphics[width=72mm]{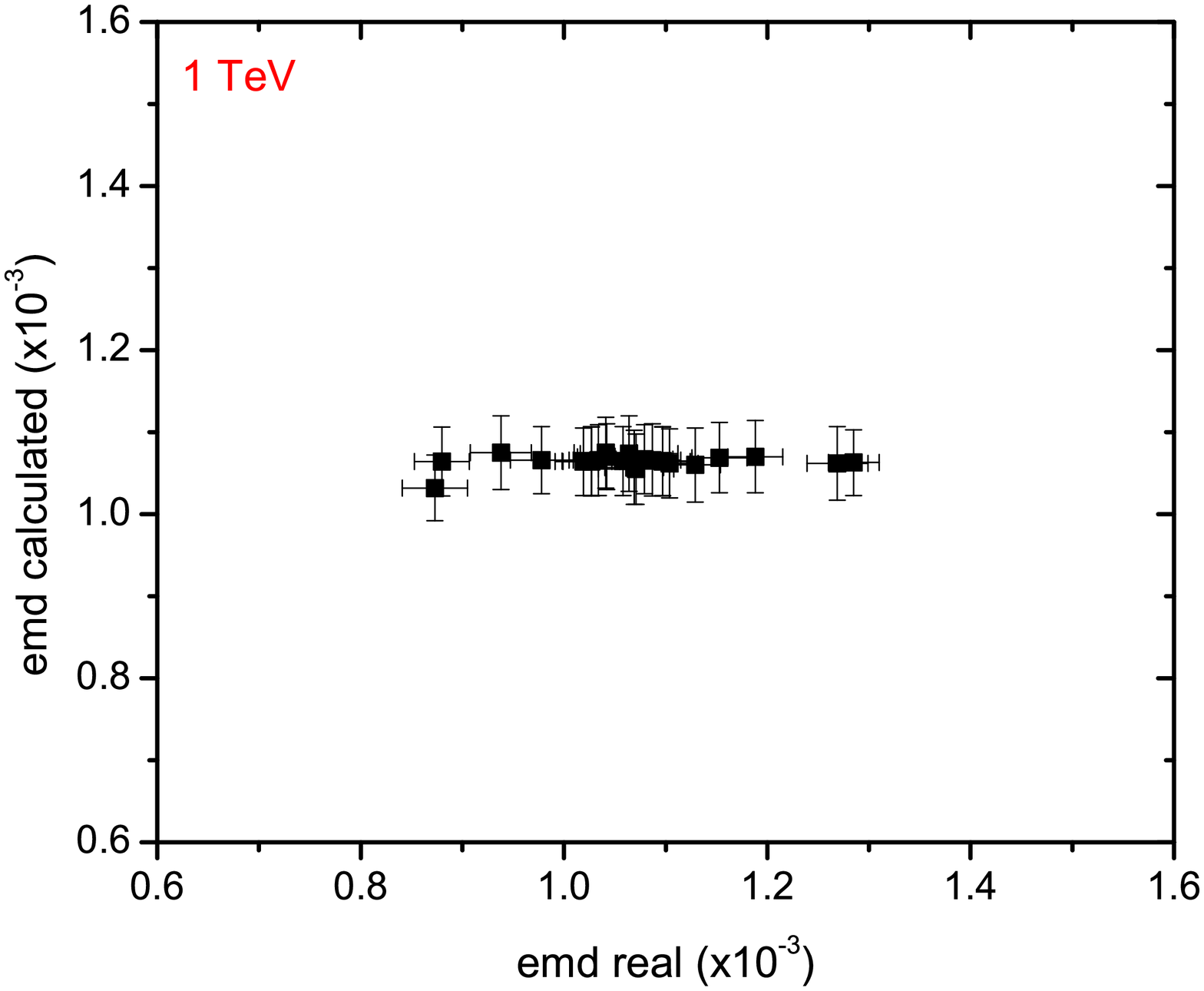}
\caption{\label{fig-emd}Scatter plot of the calculated EMD fractional counting loss obtained using the mean effective deflection angle $\Delta \theta$ derived for the nominal beam parameters, against the fractional EMD loss obtained directly from the difference in counts, for all simulated beam imperfections. The 500 GeV case is shown on the left and the 1 TeV case on the right.}
\end{figure}

The value of $\Delta \theta$ obtained in this way can be used in the experiment to determine $\Delta L_{EMD} / L$. The limitation of this correction is that it depends on simulation and on the precision of the knowledge of the beam parameters. Figure \ref{fig-emd} shows the scatter plot of the values of $\Delta L_{EMD} / L$ obtained in this way versus the values of $\Delta L_{EMD} / L$ obtained directly from the
difference in counts, for the set of beam-parameter variations described in Sec. \ref{sec-sim-bh}. The absolute value of the fractional counting error due to such beam-parameter variations is always smaller than $5 \cdotp 10^{-4}$ in the 500 GeV case and $2 \cdotp 10^{-4}$ in the 1 TeV case. If the beam parameters are known with better precision (as discussed in \cite{Grah08}), the residual uncertainty of the EMD effect on integrated luminosity will be correspondingly smaller. 

\subsection{Background from physics processes}
\label{sec:4f}

\subsubsection{Two-photon processes}
\label{two-gamma}
\begin{figure}
\centering
\includegraphics[width=40mm]{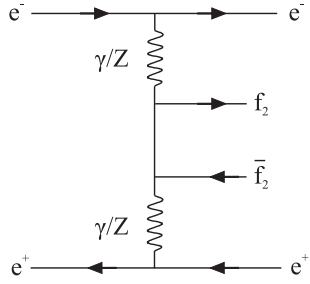}
\caption{\label{fig-Four-fermion}Two-photon exchange is the dominant Feynman diagram for neutral current four-fermion production.}
\end{figure}

Another major systematic effect in the luminosity measurement originates from the four-fermion neutral-current
processes of the type $e^{+} e^{-}\rightarrow e^{+} e^{-} f $ $\bar f$. These processes have Bhabha like signature
characterized by the outgoing $e^{+} e^{-}$ pairs emitted very close to the beam pipe, carrying a large fraction
of energy so they can be miscounted as a signal. The leading order Feynman diagram is given in Figure
\ref{fig-Four-fermion}. The dominant contribution to the cross section comes from the multiperipheral (two-photon
exchange) process.

\begin{figure}
\centering
\includegraphics{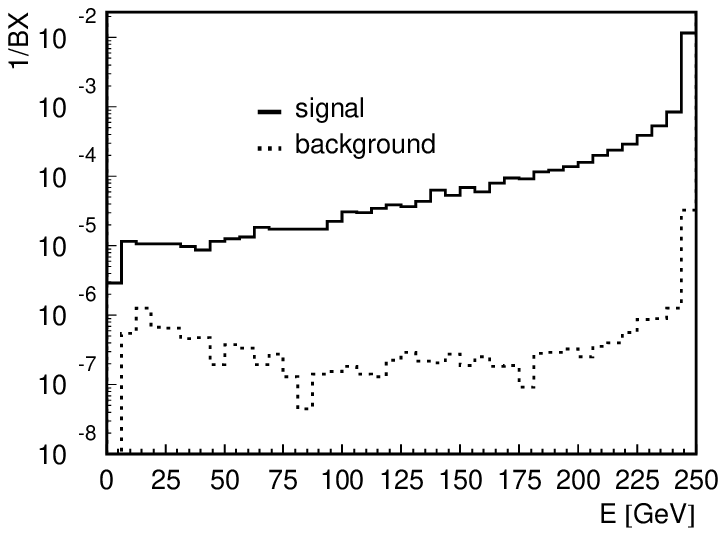}
\includegraphics{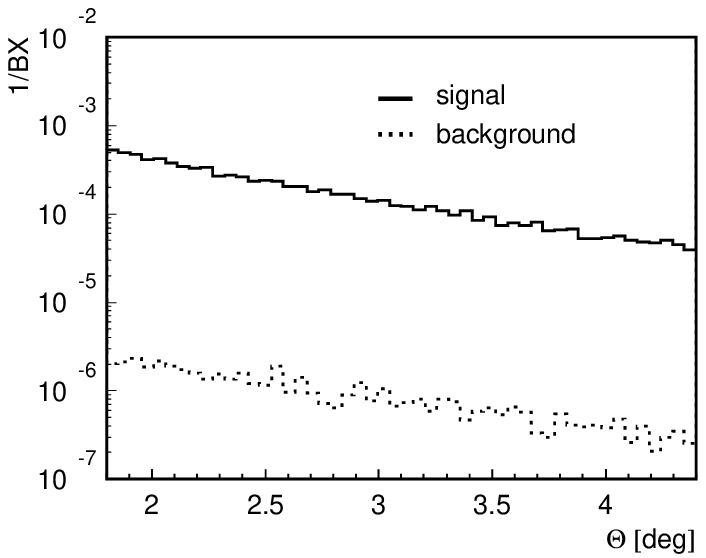}
\caption{\label{fig-energy-theta}Energy (left) and polar angle distribution (right) of final state particles originating from signal and four-fermion processes, detected in coincidence in the fiducial volume of the luminometer at 500 GeV CM energy.}
\end{figure}

Due to the steep polar angle distribution of the produced particles, only a few permille of the produced primary $e^{+} e^{-}$
pairs are deposited in the luminometer. The rest is detected in the beam calorimeter. Thus physics
background is present in  only 6.0 (2.2) permille at 500 GeV (1 TeV) of signal cases before any selection applied. Energy
and polar angle distributions of signal and physics background in the acceptance region of the luminosity calorimeter, at 500
GeV CM energy are given in Figure \ref{fig-energy-theta}. Similar distributions are obtained at 1 TeV CM energy. In Figure \ref{fig-energy-theta} (left), depositions at the nominal beam energy come from the electron spectators. Physics
background polar angle distribution is similar to the one for the signal (Figure \ref{fig-energy-theta}, right), confirming the very forward nature of the four-fermion production.

\begin{figure}
\centering
\includegraphics{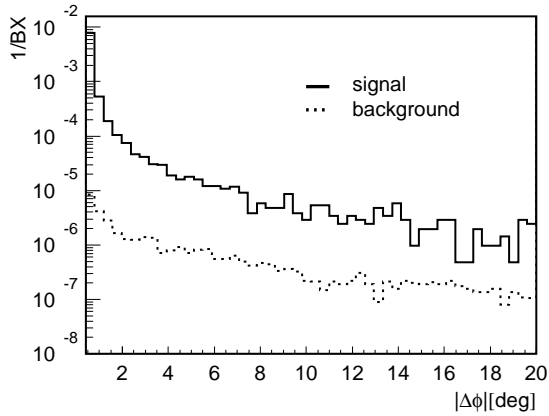}
\caption{\label{fig-delta_phi_S_B} Difference in azimuthal angle of particles detected in the fiducial volume at the opposite sides of the luminometer, for signal (solid line) and background (dashed line) at 500 GeV CM energy.}
\end{figure}	

\subsubsection{Event selection}
\label{evt-SEL}

In this paper we propose a selection that minimizes the overall impact of the leading systematic effects (beam-induced effects, background from physics processes) on the uncertainty of the integral luminosity measurement. 
Performance of the proposed selection is described by the two variables, signal efficiency $E_{s}$ and background
rejection $R_{bck}$:

\begin{equation}
\label{eq-sig-eff}
E_{s} = \frac{N^{'}_{s}}{N_{s}}\qquad\qquad
R_{bck} = 1-\frac{N^{'}_{bck}}{N_{bck}}
\end{equation}
where $N_{s}$, $N_{bck}$ correspond to the number of coincidently detected pairs at the opposite sides of
luminometer, for signal and background respectively, while the prime values correspond to the number of selected events.
Acoplanarity, given in Figure \ref{fig-delta_phi_S_B}, can be useful as a signal to background separation variable, giving the background rejection above 50\%
\cite{Mila00} for $|\Delta \phi|<5^{o}$.

In order to be compatible with the Bhabha counting method presented in Section \ref{sec:coll}, all selection cuts must be invariant with respect to the boost along the beam-beam axis. A natural selection criterion is to require that the reconstructed CM energy is higher than 80\% of the nominal CM energy, because of the heavy losses of Bhabha events below that energy (Section \ref{sec:coll}). Another selection criterion follows from the approximate coplanarity of the Bhabha events. Difference in azimuthal angle $|\Delta \phi|$ of particles detected at the opposite sides of the luminometer is given for signal and background in Figure \ref{fig-delta_phi_S_B}. As $|\Delta \phi|$ is invariant to the longitudinal boost, it can be used to suppress background. Thus, in addition to the CM energy cut, it will be required that $|\Delta \phi| < 5^{o}$. As a side effect, the criterion on acoplanarity reduces the fraction of events with $\beta_{coll} > \beta^*$ (Section \ref{sec:coll}), from approximately 1.4 permille to 0.4 permille at all ILC energies. 

Signal selection efficiency and background rejection for the proposed selection are given in Table 1, for
leptonic and hadronic background at 500 GeV and 1 TeV CM energies. Since NLO corrections for the four-fermion
production cross section are not yet available at ILC energies, instead of correcting for miscounts due to the presence of
background processes, we will assume the full-size contribution of physics background to the relative systematic uncertainty of the
measured luminosity.

\begin{table}[htp]
\caption{Selection and rejection efficiencies ($E_{s}$ and $R_{bck}$) for signal and background at 500 GeV and 1 TeV CM energies. }
\label{table:b/s}
\begin{center} \begin{tabular}{ |c|c|c|c| }
\hline
\textbf{}                      &                   &  500 GeV         &  1 TeV    		\\ \hline 

\hline \hline
Signal                                        & $E_{s}$   &  94 $\%$             &   94 $\%$          	\\ \hline
Leptonic background                           & $R_{bck}$ &  60$\%$              &   56$\%$           	\\ \cline{2-4}
$e^{+}e^{-} \rightarrow e^{+}e^{-}e^{+}e^{-}$ & $B/S$     & $1.6 \cdotp10^{-3}$  &  $0.7 \cdotp10^{-3}$ \\ \hline
Hadronic background                           & $R_{bck}$ &  70 $\%$             &   91 $\%$          	\\ \cline{2-4}
$e^{+}e^{-} \rightarrow e^{+}e^{-}q \bar q$   & $B/S$     & $0.6 \cdotp10^{-3}$  &  $0.1 \cdotp10^{-3}$ \\ \hline
\hline \hline 

\textbf{$\Delta L/L$}                         &        &\textbf{$2.2 \cdotp10^{-3}$}&\textbf{$0.8 \cdotp10^{-3}$}\\ \hline

\end{tabular} \end{center}
\end{table}

The cross section of the two-photon processes is rising with the CM energy as $~ln^{2}(s)$ \cite{LCWS10} and
practically saturates at several nb at a few hundred GeV. On the other hand, the polar angle distribution of the electrons which carry at least 80\% of beam energy is shifted toward lower polar angles. Thus the
majority of the $e^{+} e^{-}$ spectators from physics background are missing the luminometer.
This change in topology compensates the slight rise of the cross section, resulting in improved B/S ratio at 1 TeV
in comparison to the 500 GeV case, as visible in Table \ref{table:b/s}.

At all ILC energies, signal selection efficiency is maintained above 90\%,  while 56\% to 91\% of background is suppressed at
different CM energies depending on the type of four-fermion process. The above corresponds to the systematic
uncertainty of the integrated luminosity originating from physics  background at the level of few permille or better, while the
statistical error of the luminosity measured over a year of ILC  operation is kept at the level of $10^{-4}$ due to the high
signal efficiency.

\subsection{Summary on systematic uncertainties in luminosity measurement}
\label{syst_sum}
Various sources of systematic uncertainty that have impact on luminosity measurement are discussed in detail in
\cite{JINST2010} for the 500 GeV CM energy ILC. We take all the sources of systematic uncertainty as in
\cite{JINST2010}, except for the beam-induced effects and physics background that are revised in this paper. We extend the overview of
systematic uncertainties to the 1 TeV CM ILC case. New results also include not only pinch effect and
beamstrahlung, but also initial state radiation and electromagnetic deflection both at 500 GeV and 1 TeV (Table \ref{suTABLE}).
Differently from \cite{JINST2010}, physics background is taken in its full size without assumptions on the cross-section
uncertainty induced by NLO corrections at ILC energies. 

\begin{table}[htp]
\caption{Summary on systematic uncertainties in luminosity measurement at ILC, without$^{1}$ or
with$^{2}$ simulation-dependent corrections.}
\label{suTABLE}
\begin{center} \begin{tabular}{ |c|c|c| }
\hline
Source of uncertainty  &  $\Delta L/L$ (500 GeV)  &  $\Delta L/L$ (1 TeV)  \\ \hline 

\hline \hline

Bhabha cross-section $\sigma_{B}$  &  $5.4 \cdotp10^{-4}$  &  $5.4 \cdotp10^{-4}$  \\ \hline
Polar angle resolution $\sigma_{\theta}$  &  $1.6 \cdotp10^{-4}$  &  $1.6 \cdotp10^{-4}$  \\ \hline
Bias of polar angle $\Delta \theta$  &  $1.6 \cdotp10^{-4}$  &  $1.6 \cdotp10^{-4}$  \\ \hline
IP lateral position uncertainty  &  $1 \cdotp10^{-4}$  &  $1 \cdotp10^{-4}$  \\ \hline
Energy resolution $a_{res}$  &  $1.0 \cdotp10^{-4}$  & $1.0 \cdotp10^{-4}$  \\ \hline
Energy scale  &  $1.0 \cdotp10^{-3}$  & $1.0 \cdotp10^{-3}$  \\ \hline
Beam polarization  &  $1.9 \cdotp10^{-4}$  & $1.9 \cdotp10^{-4}$  \\ \hline
Physics background B/S  &  $2.2 \cdotp10^{-3}$  &  $0.8 \cdotp10^{-3}$  \\ \hline
Beamstrahlung + ISR$^{1}$  &  $-1.1 \cdotp10^{-3}$  &  $-0.7 \cdotp10^{-3}$  \\ \hline
Beamstrahlung + ISR$^{2}$  &  $0.4 \cdotp10^{-3}$  &  $0.7 \cdotp10^{-3}$  \\ \hline
EMD$^{1}$  &  $-2.4 \cdotp10^{-3}$  &  $-1.1 \cdotp10^{-3}$  \\ \hline
EMD$^{2}$  &  $0.5 \cdotp10^{-3}$  &  $0.2 \cdotp10^{-3}$  \\ \hline

\hline \hline

$(\Delta L/L)^{1}$  &  $4.3 \cdotp10^{-3}$  &  $2.3 \cdotp10^{-3}$  \\ \hline
$(\Delta L/L)^{2}$  &  $2.6 \cdotp10^{-3}$  &  $1.6 \cdotp10^{-3}$  \\ \hline

\end{tabular} \end{center}
\end{table}

Uncertainties are assumed to be uncorrelated. The only exception is for the remaining biases due to beamstrahlung and EMD in the
case of simulation-independent correction. Since these biases are both negative by definition, they were linearly summed
together and that sum is then added quadratically to the uncertainties from other sources. 
Uncertainty of the theoretical cross-section for Bhabha scattering is taken to be as at LEP energies
\cite{Jad03}. As can be seen from Table \ref{suTABLE}, overall systematic uncertainty of
luminosity is no larger than a few permille.

\section{Summary}

Interaction of the colliding beams introduce sizeable bias of the order of 10 percent in luminosity measurement at ILC
energies. This dominantly comes from the asymmetric energy loss of the initial state due to the beamstrahlung and ISR. Existing
corrective techniques are simulation dependent and the systematic uncertainty of luminosity measurement depends on the
knowledge of the beam parameters. Within the proposed selection, a method to correct for these effects in a simulation independent manner and
practically without sensitivity to the knowledge of the beam parameters is presented in this paper. It is demonstrated
that the systematic bias induced by the beamstrahlung and ISR can be reduced by a factor of a hundred to the level of
1.1 permille or better at ILC energies.

If simulation is used to correct the off-axis ISR contribution, the uncertainty due to the beamstrahlung and the ISR effects drops below one permille. In addition, simulation dependent corrections can be applied to reduce the already small effect (of order of a permille) of the electromagnetic deflection of the final state. The overall residual uncertainty of luminosity measurement induced by the beam-beam effects is in this case below permille. 

Dominant source of uncertainty in luminosity measurement comes from the physics
background taken as a full-size effect until reliable estimates of the cross-section for the four-fermion production at ILC energies become available. 

Taking into account all the sources of systematic uncertainty of luminosity measurement we show that the luminosity can
be measured at ILC in the peak region above 80\% of the nominal CM energy with the required precision of a few
permille, even if one corrects for the dominant systematic effects without using simulation-dependent information.            

\acknowledgments

We express our gratitude to our colleagues within the FCAL and CLIC collaborations for constructive reviewing of the
presented methods. We also acknowledge the support received from the Ministry of education, science and technological
development of the Republic of Serbia within the project OI171012.
 
\bibliographystyle{JHEP}
\bibliography{lumi}

\end{document}